\documentclass[iop,apj,numberedappendix,twocolappendix,revtex4]{emulateapj}
\usepackage{epstopdf}

\usepackage{multirow}

\slugcomment{}

\shorttitle{Vertical structure of ADAFs}
\shortauthors{Zeraatgari \& Abbassi}

\begin{document}
\title
{Vertical structure of Advection dominated Accretion Flows}

\author {Fateme Zahra Zeraatgari\altaffilmark{1}and Shahram Abbassi\altaffilmark{1,2}}

 \altaffiltext{1}{Department of Physics, School of Sciences, Ferdowsi University of Mashhad, Mashhad, 91775-1436, Iran; fzeraatgari@yahoo.com, abbassi@um.ac.ir}
 \altaffiltext{2}{School of Astronomy, Institute for Research in Fundamental Sciences (IPM), Tehran, 19395-5531, Iran}

\begin{abstract}
We solve the set of hydrodynamic (HD) equations for optically thin Advection Dominated Accretion Flows (ADAFs) by assuming radially self-similar in spherical coordinate system $ (r, \theta, \phi) $. The disk is considered to be steady state and axi-symmetric. We define the boundary conditions at the pole and the equator of the disk and to avoid singularity at the rotation axis, the disk is taken to be symmetric with respect to this axis. Moreover, only the $ \tau_{r \phi} $ component of viscous stress tensor is assumed and we have set $ v_{\theta} = 0 $.  The main purpose of this study is to investigate the variation of dynamical quantities of the flow in the vertical direction by finding an analytical solution. As a consequence, we found that the advection parameter, $ f^{adv} $, varies along the $ \theta $ direction and reaches to its maximum near the rotation axis. Our results also show that, in terms of no-outflow solution, thermal equilibrium still exists and consequently advection cooling can balance viscous heating.

\end{abstract}


\section{INTRODUCTION}

Accretion onto a compact object such as black hole is a fundamental phenomenon in the universe and most likely is the
primary power source in systems like X-ray binaries (XRBs), active galactic nuclei (AGNs) and gamma-ray bursts (GRBs).
Observation of X-ray and gamma-ray emission lines from black hole accretion disks demonstrates the idea of forming a hot
atmosphere above the accretion disk or perhaps radiatively inefficient flows (RIAFs) (e.g. Blair et. al. 1984; Raymond 1993; Jimenez-Garate et. al. 2005.)

In the standard accretion disk model (Shakura \& Sunyaev 1973) the energy released via viscosity is radiated locally
and the accreting flow becomes cool very efficiently. Therefore, this model cannot produce high energy spectrum and
the idea of existence of hot corona above disk is needed to predict such high-energy emissions.
In terms of RIAFs, in fact, the generated heat through viscosity is stored as entropy and can be transported
with flow inwardly rather than immediately being radiated away from the system. Consequently, the flow
temperature becomes extremely high and nearly the virial temperature. As a result, the disk can radiate high-energy
emission such as gamma-ray (see Kato et al. 2008; Yuan \& Narayan 2014 for more details).
The disks and flows with this essential feature are named optically thin advection-dominated accretion flows (ADAFs).
Historically, the significance of advection energy in hot accretion flows was firstly recognized by Ichimaru 1977
and more importantly, a wide range of studies on optically thin ADAFs has been done by Narayan \& Yi 1994,
Narayan \& Yi 1995a,b and Abramowicz et al. 1995 where the disk was thermally stable (more details in Yuan \& Narayan 2014).

It should also be mentioned that in the field of hot accretion flows, a lot of numerical HD and MHD simulations have been carried out to investigate the dynamics of hot accretion flows and one of the most important findings by those simulations is that the mass inflow rate decreases inwardly (e.g., Igumenshchev \& Abramowicz 1999, 2000; Stone, Pringle \& Begelman 1999; Hawley, Balbus \& Stone 2001; De Villiers, Hawley \& Krolik 2003; Igumenshchev, Narayan \& Abramowicz 2003; Yuan \& Bu 2010; Pang et al. 2011; Yuan et al. 2012a,b; Bu et al. 2013) and it is not constant as considered before. Following those simulation results, one and two dimensional self-similar solutions of ADAfs in the presence of outflow and magnetic field have been done (e.g., Xu \& Chen 1997; Blandford \& Begelman 1999, 2004; Xue \& Wang 2005; Akizuki \& Fukue 2006; Abbassi et al. 2008; Zhang \& Dai 2008; Bu et al. 2009; Jiao \& Wu 2011, Abbassi \& Mosallanezhad 2013; Mosallanezhad et al. 2013, 2014, Samadi et al. 2015). Note that, the properties and dynamic of the hot accretion flows with magnetic field and outflow is beyond the scope of this work.

In the optically thin ADAFs, the accretion rate is very low, $ \dot{M} \lesssim 0.1 L_{Edd}/c^2$ where $ L_{Edd} $ is the Eddington luminosity and $ c $ is the speed of light. In addition, the optically thin ADAFs are geometrically thick disks, i.e., $ H/R \lesssim 1$, where $ H $ is the disk's height scale and $ R $ is the radius in cylindrical coordinates. It should be noted here that in the novel vertically average self-similar methodology presented by Narayan \& Yi 1994 (hereafter NY94), the energy equation expressed as,
\begin{equation} \label{NY energy}
	q^{adv} = q^{+} - q^{-} \equiv f q^{+}
\end{equation}
where $ q^{adv} $ represents the rate of the entropy advection in the radial direction, i.e., $ \rho v_{r} T ds/dR $. Where, $ \rho $ is
the density of the gas at the equatorial plane of the disk,  $ v_{r} $ is the radial velocity and also $ s $ and $ T $ are the
specific entropy and temperature of the gas respectively. In the equation (\ref{NY energy}), $ q^{+} $ gives the total heat
generated by viscosity per unit volume per unit time in the radial direction. They defined the advection parameter as
$ f \equiv q^{adv}/q^{+}$ which measures the fraction of the advection energy stored as entropy. Consequently,
$ (1 - f) $ will be radiated away from the system. They integrated the flow equation in the vertical direction. Making usual assumption such as steady state, axisymmety and $\alpha$-viscosity they obtained a set of ordinary differential equations for the variables as a function of $r$. They have shown that the equations has a exact self-similar solution where all variables have power-law dependencies on $r$. Vertical integration is an standard approximation which has been used for thin disks where vertical thickness is usually much smaller than the local radius. While as we mentioned, optically thin ADAFs are geometrically thick disks,
therefore, height-integrated approximation is not appropriate because the physical variables are not only function of $ r $ but also
they should be function of the vertical direction, $ \theta $. Then, in the case of optically thin ADAFs, 1D approach is not
suitable.  Later on, Narayan \& Yi 1995a (hereafter NY95a), tried to solve the Hydrodynamic equations (HD) in spherical polar coordinates $ (r, \theta) $ with
radially self-similar solutions but their results only corresponded to the simplest form of the advection parameter, viz., $ f = constant $.

 To answer how the advection parameter varies along spherical polar angle, Gu et al. 2009 adopted a polytropic relation,
 $ p = K {\rho}^{\Gamma} $, in the vertical direction which is normally used in vertically integrated geometrically thick disk model
 (e.g., Kato et al 2008). They defined the inclination $ \theta_{s} $ near the polar axis regarding as the surface of the disk.
 Therefore, the main conclusion was that the optically thin ADAFs are geometrically thick since the free surface of the disk is very close
 to the polar axis. By taking into account the effect of a toroidal magnetic field and its corresponding heating, Samadi et al. 2013
determined the thickness of advection-dominated accretion flows. Their results show that the vertical component of magnetic force acts in the opposite direction to gravity and
compresses the disc; thus, compared with the non-magnetic case, in general the disc half-thickness,
$\Delta\theta$, is significantly reduced. It should be emphasized that
 in both above mentioned works, the power index $ \Gamma $ considered with a typical value above unity, for example,
 $  \Gamma = 4/3 $. Also, the constant $ K $ is set equal to one, $ K = 1.0 $, to solve their one boundary differential equations starting from surface of the disk.
 Simulations carried out by De Villiers et al. 2005 revealed that the time average density drops faster than pressure in the
vertical direction which means that the power index $ \Gamma $ should be less than one to satisfy the polytropic relation
 in the $ \theta $ direction.

A few analytical solutions in the case of hot accretion have been presented and such solutions need to assume some simplifications. For instance, Shadmehri 2014 considerd the energy equation of the gas with the inclusion of only $ \tau_{r\phi} $ component of viscosity in the viscous energy dissipation term and found a very creative analytical solution. Later on, Based upon new simulation results mentioned above,  Gu 2015  (hereafter G15) repeated his previous work, Gu 2009, with an original idea of finding an analytical solution. The behavior of physical variables are in satisfactory with those in NY95a except for the variation of isothermal sound speed and radial velocity profiles along the vertical direction. The main conclusion was, viscous heating and advection cooling cannot balance each other. Therefore, no thermal equilibrium exists under the purely inflow assumption.

 The analytical solution we present here is in the same methodology as described in G15, with three modifications.
 Firstly, we define the first boundary conditions at the rotation axis $ \theta = 0^{\circ} $ to increase the angular range of
 our calculation. we made this change because in terms of optically thin ADAFs, the disk is considered to be geometrically
 thick and there might exist low-dense with high temperature flow above the surface of the disk.
 Secondly, following NY95a, to avoid singularity at the poles, the disk is taken to be symmetric with respect to this axis.
 The second change leads to find a relation between the value of constant $ K $ and the density at the rotation axis.
 Finally, we will adopt the modified $ " \alpha " $ description of viscosity defined by Bisnovatyi-Kogan \& Lovelace 2007. In the next section we explain with more details why this form of viscosity is needed. With the aforementioned modification, we can address whether there exist thermal equilibrium in the purely inflow case and check how the advection parameter change
 along the vertical direction.

The outline of this paper is as follows. In section 2, the basic equations and boundary conditions are introduced. The numerical results are shown and discussed in more details in section 3. Finally, a brief summery and conclusions will be given in section 4.

\section{BASIC EQUATIONS AND BOUNDARY CONDITIONS}
\subsection{Basic equations}

The standard hydrodynamic (HD) equations are employed in spherical coordinate system $ (r, \theta, \phi) $
where the steady state accretion flow is taken to be axisymmetric (i.e., $ \partial/\partial \phi  = 0 $).
The gravitational potential of the central black hole is described in terms of Newtonian potential which is
more convenient for the self-similar formalization, $ \psi(r) = - (GM)/r $. In addition, the flow is in
non self-gravitating regime and initially the relativistic effects are neglected. Following NY95a,
we assume $ v_{\theta} = 0 $ , which corresponds to a hydrostatic equilibrium in the vertical direction.
Although, this assumption is not appropriate if you are investigating the effects of outflow on the dynamics
of accretion flow (see e.g., Jiao \& wu 2011; Mosallanezhad et al. 2014 for more details).
Therefore, the continuity equation and the three components of the equation of motion are as follows,
\begin{equation}\label{continuity}
  \frac{1}{r^{2}} \frac{\partial}{\partial r} \left( r^{2} \rho v_{r} \right) = 0,
\end{equation}
\begin{equation}\label{motion_r}
	v_{r} \frac{\partial v_{r}}{\partial r} - \frac{v_{\phi}^{2}}{r} = - \frac{GM}{r^{2}} - \frac{1}{\rho} \frac{\partial p}{\partial r},
\end{equation}
\begin{equation}\label{motion_theta}
	v_{\phi}^{2} \cot \theta = \frac{1}{\rho} \frac{\partial p}{\partial \theta},
\end{equation}
\begin{equation}\label{motion_phi}
	v_{r} \frac{\partial v_{\phi}}{\partial r} + \frac{v_{r} v_{\phi}}{r} = \frac{1}{\rho r^{3}} \frac{\partial}{\partial r} \left( r^{3} \tau_{r \phi} \right).
\end{equation}
where $ v_r $ and $v_{\phi} $ are radial and azimuthal components of velocity,
$ \rho $ is the mass density and $ p $ stands for the gas pressure. Besides, in equation (\ref{motion_phi}),
$ \tau_{r \phi} $ represents the $ r \phi $ component of the anomalous stress tensor, respectively.
It should be emphasized that in a real case, the magnetic stress driven by the magneto-rotational
instability (MRI) transfers the angular momentum outside the disk (Balbus \& Hawley 1991, 1998).
Since in our HD case we do not consider magnetic field, the anomalous shear stress tensor
has been considered to mimic the magnetic stress (see the HD simulations performed by Yuan et al. 2012a
for more details). This parameter can be written as,
\begin{equation}\label{tau_rphi}
  \tau_{r \phi} = \mu r \frac{\partial}{\partial r} \left( \frac{v_{\phi}}{r} \right).
\end{equation}
where $ \mu (\equiv \nu \rho) $ is the viscosity coefficient which determines the magnitude of the stress
and $ \nu $ is called the kinematic viscosity coefficient. There are a lot of uncertainties about how to
prescribe such a  viscosity parameter. Most researchers adopt the $ `` \alpha " $ description
for standard thin disks introduced by Shakura \& Sunyaev 1973 which is proportion to speed of sound as
$  \nu = \alpha h c_{s} $. Here, $ h $ is the disk's height scale and $ \alpha $  is a constant parameter less
than unity. We know that, if the viscosity coefficient scales with radius as $ \nu \propto r^{1/2} $ then the
radial self-similarity will be possible.

It should be point out that some simulations have been carried out for different
forms of viscosity coefficients in accretion disks which found that the azimuthal
components dominate other components (e.g. Stone et al. 1996), therefore, in
HD calculations, it would be more convenient to take into account the azimuthal
components (see, Stone et al. 1999; Yuan et al. 2012a,b).

NY95a obtained their solutions for  ADAFs model corresponding to the usual $ `` \alpha " $ description  of viscosity.
They also checked whether the results are sensitive to the viscosity by adopting different form as
$ \nu = \alpha r c_{s} $ and concluded that their solutions with new form are entirely similar to those with
$ `` \alpha " $ description.
This prescription may not be suitable in a real case, since there exists a low-dense corona above the disk with a nearly
virial temperature. In addition, in the case of geometrically thick and hot disk (ADAFs model) the hottest temperature
should be achieved at the rotation axis, $ \theta = 0, \pi $ (see Figure 1 of NY95a for more details).
On the other hand, the viscosity is due to MRI turbulence and this quantity should vanish at the surface of the disk
(Bisnovatyi-Kogan \& Lovelace 2007; Lovelace et al. 2009). Then, if the kinematic viscosity coefficient is proportional to the isothermal sound speed, $ \nu \propto c_{s} $, then this quantity cannot vanish at the surface of the disk as MRI turbulence predicts.

In order to avoid the disparateness in terms of the turbulent viscosity, following Lovelace et al. 2009, we adopt the modified $ " \alpha " $ description of viscosity as
\begin{equation}\label{nu_description}
  \nu = \alpha \frac{c_{s}^{2}}{\Omega_{k}} g \left(\theta\right)
\end{equation}
In the above equation, $ \Omega_{K} (\equiv \sqrt{GM/r^{3}}) $ is the Keplerian angular velocity of the disk and
$ g \left(\theta\right) $ is a dimensionless function which is equal to unity and zero in
the body and surface of the disk, respectively (e.g., Lovelace et al. 2009).
For simplicity, we consider $ g \left(\theta\right) = \sin \theta $ to satisfy the aforementioned conditions.

We are interested in investigating whether the advection parameter, $ f $, which is normally considered to unity
$ f = 1 $, in the case of ADAF, remains constant along the polar angle or not (e.g., Narayan et al. 1995a;
Xu \& Chen 1997; Jiao \& Wu 2011). Therefore, following Gu et al. 2009 and also Gu15, we apply polytropic
relation, $ p = K \rho^{\Gamma} $ in the $ \theta $ direction as our last equation.
Although, they obtained solutions by fixing $ K = 1 $ , we explain how this constant parameter
will be determined in the next section. We also note that, the simulations of De Villiers et al. 2005 revealed
that the power index $ \Gamma $ is less than unity. This is mainly important because their results show that the time
averaged density drops faster than pressure from the equatorial plane to the polar axis. Based upon those results,
$ \Gamma $ is set to be less than one throughout this paper.

We adopt self-similar solutions in the radial direction to simplify the equations as

\begin{equation}\label{rho_selfsimilar}
  \rho(r, \theta) = \rho(\theta) r^{-3/2},
\end{equation}
\begin{equation}\label{vr_selfsimilar}
  v_{r} (r, \theta) = \sqrt{\frac{GM}{r}} v_{r}(\theta) = v_{K}(r)v_{r}(\theta),
\end{equation}
\begin{equation}\label{vphi_selfsimilar}
  v_{\phi} (r, \theta) = v_{K}(r) v_{\phi}(\theta),
\end{equation}
\begin{equation}\label{cs_selfsimilar}
   p (r, \theta) = p(\theta) GM r^{-5/2}.
\end{equation}

By substituting above self-similar solutions into Equations (\ref{continuity})-(\ref{motion_phi}), they will be reduced to
\begin{equation}\label{motion_1}
  -\frac{1}{2} {v_{r}(\theta)}^{2} - {v_{\phi}(\theta)}^{2} = -1 + \frac{5}{2}K {\rho(\theta)}^{\Gamma - 1},
\end{equation}
\begin{equation}\label{motion_2}
  {v_{\phi}(\theta)}^{2} \cot\theta  = K \Gamma {\rho(\theta)}^{\Gamma - 2} \frac{d\rho(\theta)}{d\theta},
\end{equation}
\begin{equation}\label{motion_3}
  v_{r}(\theta)= - \frac{3}{2}  \alpha  K {\rho(\theta)}^{\Gamma - 1} \sin\theta.
\end{equation}
We will put Equations (\ref{motion_1}) and (\ref{motion_3}) into Equation (\ref{motion_2}) in order to obtain the differential
equation for the density. Before doing that, it should be note that, the first term on the left hand side of Equation (\ref{motion_1})
is very small compare to the other terms. This is because in the case of ADAF, radial velocity is very low and also the viscosity
constant considered here is fixed as $ \alpha = 0.1 $ (see equation (\ref{motion_3})). Hence, without any significant change in
our results we can neglect this term and therefore, the differential equation will be written as (see G15 for more details):
\begin{equation}\label{dif_rho1}
  \frac{d\rho(\theta)}{d\theta} = \frac{\cot\theta}{\Gamma} \left( \frac{{\rho(\theta)}^{2 - \Gamma}}{K} - \frac{5}{2} \rho(\theta) \right).
\end{equation}
The above differential equation has an analytical solution which will be obtained after introducing boundary conditions in the following section.
\subsection{BOUNDARY CONDITIONS}
\begin{figure}
\begin{center}
\includegraphics[width=90mm, angle=0]{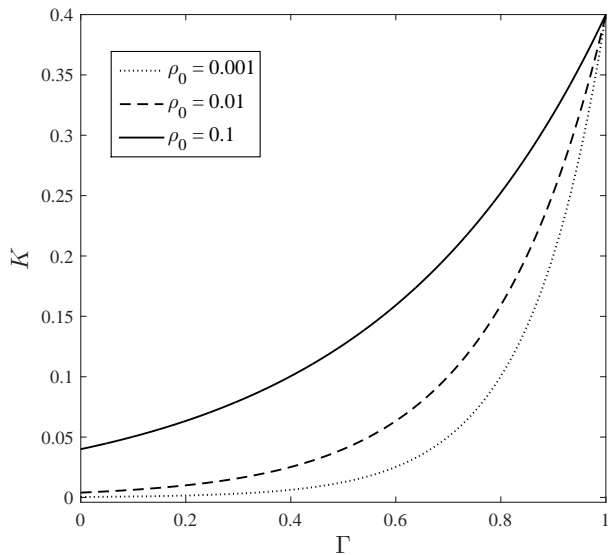}
\caption{Variation of $ K $ with $ \Gamma $ for different value of density at polar axis, $ \rho = 0.001, 0.01, 0.1 $.}
\label{K_Gamma}
\end{center}
\end{figure}

We define the boundary conditions
at the rotation axis $ \theta = 0 $, and equatorial plane, $ \theta = \pi/2 $ to occupy the angular range
$ 0 \leq \theta \leq \pi/2 $. Following NY95a, to avoid singularity at the polar axis, we assume the boundary condition as
\begin{equation}\label{boundary_0}
  \theta = 0:\qquad \frac{d\rho}{d\theta} = 0.
\end{equation}
The above boundary condition leads to obtain the value of constant $ K $. Therefore, in Equation (\ref{dif_rho1}), the term inside
the parentheses should become zero to satisfy above boundary condition. So, the $ K $  parameter can be derived as,
\begin{equation}\label{K_parameter}
  K = \left( \frac{2}{5} \right) \rho_{0}^{1  - \Gamma}
\end{equation}
where $ \rho_{0} $ represents the value of density at the rotation axis. Fig. \ref{K_Gamma} shows variation of constant
$ K $ versus $ \Gamma  $ corresponding to different values of $ \rho_{0} $. Note that, since we will fix the value of
the density at the equatorial plane to unity, $ \rho (\pi/2) = 1.0 $, the magnitude of $ \rho_{0} $ in our Figures
represents the ratio of the density at the polar axis to the mid-plane of the disk.
It is clear from equation (\ref{K_parameter}) and also Fig. \ref{K_Gamma} that $ K $ increases with increasing power
index $ \Gamma $ for different fixed values of $ \rho_{0} $. Another feature that can be seen in this figure is that
the value of $ K $ cannot exceed $ 2/5 $ which is the higher limit of this parameter with $ \Gamma $ index in a range
$ 0 < \Gamma < 1.0 $. As we explained before, we consider this range for $ \Gamma $ because according to the simulation
results carried out by De Villiers et al. 2005 the time average density drops faster than pressure in the $ \theta $ direction.
This figure is compared with Figure (2) in G15 paper because at first, we decided to modify G15. Comparing
Equation (\ref{K_parameter}) with Equation (10) inside G15 paper, obviously both will be equal if one considers
the surface angle to be $ 0^{\circ} $. In order to avoid defining the parameter $ \lambda(\equiv(p_{0}/\rho_{0})/v_{K}^2) $ as the energy advection on the midplane of the disk (G15), we instead use the constant $ K $ which is the coefficient inside the polytropic equation.

Now, we turn our attention to find analytical solution for the density in the vertical direction. By integrating Equation
(\ref{dif_rho1}) along the $ \theta $ direction, we can easily obtain the density profile as
\begin{equation}\label{analytical solution}
  \rho(\theta) = \left\{ \frac{1}{5K}\left[  \left( 5K {\rho(\pi/2)}^{\Gamma - 1} - 2 \right) \sin\theta^{\frac{-5(\Gamma - 1)}{2\Gamma}}  + 2 \right]  \right\}^{\frac{1}{\Gamma - 1}}
\end{equation}
where the value of density at the equatorial plane is set to be  $ \rho(\pi/2) = 1.0 $ throughout this paper.
To complete the specification of the results, we need to define the advection parameter, $ f^{adv} $.
In the self-similar formalism, the advective cooling rate and the viscous heating rate per unit volume can be expressed as
\begin{equation} \label{advection cooling}
	q^{adv} = - \frac{5 - 3 \gamma}{2\left( \gamma - 1 \right)} \frac{p v_{r}}{r},
\end{equation}	
\begin{equation}\label{viscous heating}	
	q^{+} = \frac{9 \alpha}{4} \frac{p v_{\phi}^{2}}{r v_{K}} sin\theta.
\end{equation}
Therefore, by the vertical integration over $ q^{adv} $ and $ q^{+} $, we can achieve $ Q^{adv} $ and also $ Q^{+} $ as
\begin{equation}\label{Q_adv}
	Q^{adv} = 2 \int_{0}^{\frac{\pi}{2}} q^{adv} r \sin\theta\ d\theta
\end{equation}	
\begin{equation}\label{Q_vis}
    Q^{+} = 2 \int_{0}^{\frac{\pi}{2}} q^{+} r \sin\theta\  d\theta
\end{equation}
Then, the energy advection factor is given by $ f^{adv} \equiv Q^{adv}/Q^{+} $. In the next section we will
express the behavior of all variables and also the variation of advection cooling and viscous heating by
explanation and comparing our results to those in Gu15 and Gu 2009.
\begin{figure*}
\begin{center}
\includegraphics[width=\textwidth, angle=0]{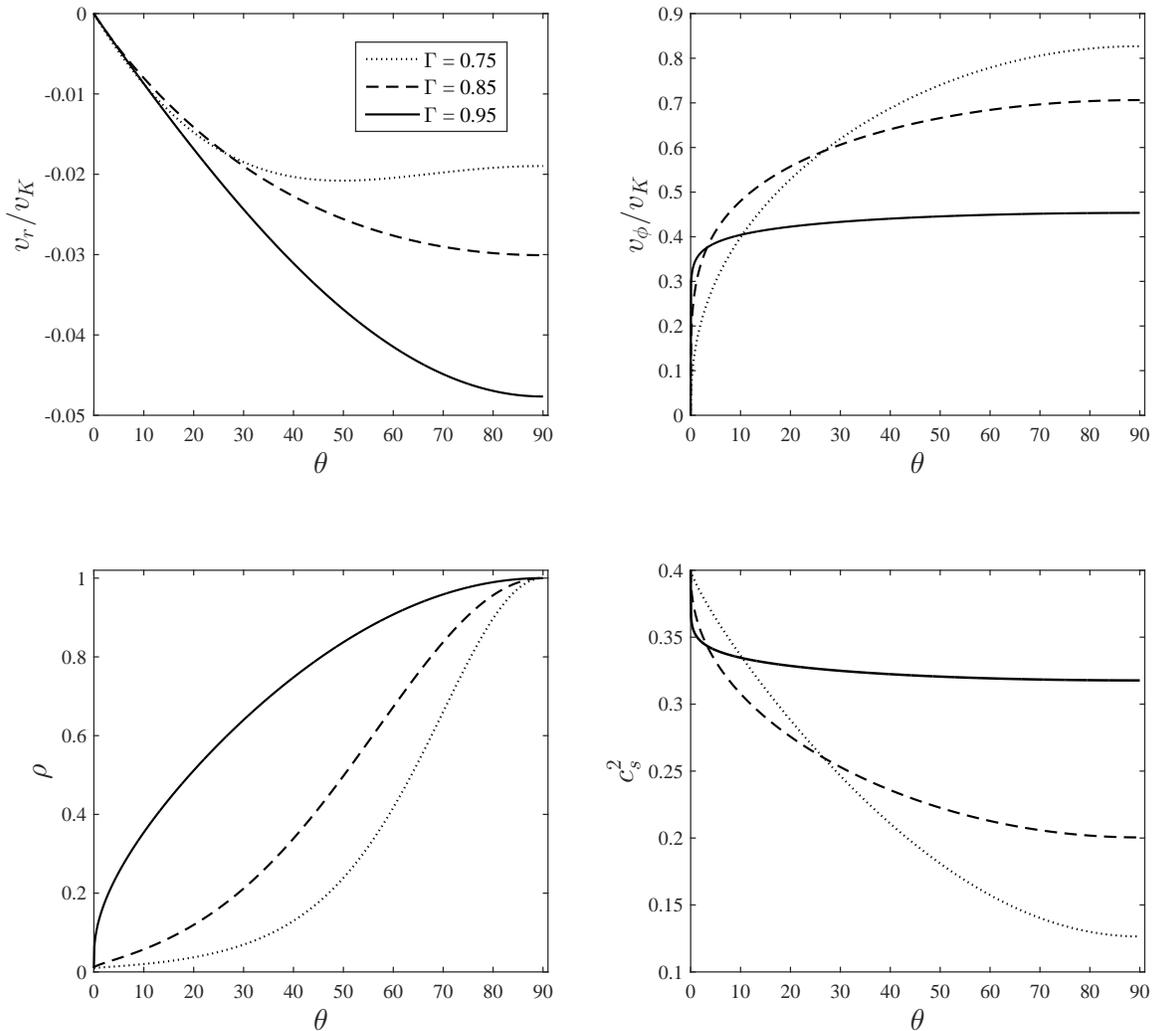}
\caption{Variation of dimensionless physical quantities with polar angle $ \theta $. The dotted, dashed and solid lines correspond to $ \Gamma = 0.75, 0.85, 0.95 $ respectively. Here $ \alpha = 0.1 $ and $ \rho_{0} = 0.01 $. }
\label{variables}
\end{center}
\end{figure*}
\section{numerical results and discussion}
In this section, we first study the angular distribution of the dynamical quantities driven from Equations
(\ref{motion_1})-(\ref{motion_3}) and Equation (\ref{analytical solution}) as well. Fig. \ref{variables}
shows the variation of physical quantities with the polar angle, corresponding to $ \Gamma = 0.75, 0.85, 0.95 $ as
three typical examples and the density at the polar axis as $ \rho_{0} = 0.01 $.
The dimensionless radial velocity, $ v_{r}/v_{K} $, is plotted in the top left panel of Fig. \ref{variables}. As it is seen,
the radial velocity increases from the rotation axis to the equatorial plane. Actually, $ v_{r} $ is zero at $ \theta = 0^{\circ} $
and reaches to it's maximum at $ \theta = \pi /2 $ for all $ \Gamma $ power index values. It should be note here that, in terms of
$ \Gamma < 1$, G15 found that the radial velocity increases toward the rotation axis (see Fig. 1, panel (b) of G15). This contradiction
is simply because, in this paper the modified form of $ \alpha $ prescription for viscosity is adopted (see Equation
(\ref{nu_description}) and also Equation (\ref{motion_3}) for more details). In addition, our result shows that the larger $ \Gamma $
leads $ v_{r} $ increases further.

The top right panel displays the dimensionless azimuthal velocity, i.e., $ v_{\phi }/v_{K} $. It is clear that $ v_{\phi } $ increases
from $ \theta = 0^{\circ} $ to $ \theta = \pi /2 $ and very close to the pole, $ v_{\phi} $ vanishes and becomes zero. This behavior
is clearly because, as a second modification to G15, the disk is taken to be symmetric with respect to the polar axis. As a result,
the centrifugal force will become zero at the rotation axis. Moreover, azimuthal component of velocity is larger for small values of $ \Gamma $.
In fact, the most variation of $ v_{\phi } $ is belonged to the smaller value of $ \Gamma $. As seen in this panel $ v_{\phi} $ is
changed from $ v_{\phi} = 0 $ to just above $ v_{\phi} = 0.8 $ for $ \Gamma = 0.75 $.

The bottom left panel shows the vertical profile of the density $ \rho $. It should be emphasized that the density profile is scale with the density
value on the equatorial plane of the disk. In this figure, the minimum value of the density at the rotation axis is considered as $ \rho_{0} = 0.01 $.
As an overall trend, it is clear that the density increases from $ \rho ‎\simeq‎ 0 $ at the rotation axis to $ \rho = 1 $ at the equatorial plane.
What's more, for the larger $ \Gamma $ as you can see, $ \rho $ increases sharper than the smaller one which means there exists extremely dense flow near the rotation axis and the disk is considered to be geometrically thick.

Finally, the bottom right panel of Fig. \ref{variables} shows the vertical variation of the isothermal sound speed, $ c_{s}^{2} $.
As illustrated in this panel, $ c_{s}^{2} $ has the decreasing trend from the rotation axis to the equatorial plane. Furthermore, from the figure it is clear that for the case
with $ \Gamma = 0.95 $, $ c_{s}^{2} $ is almost independent of $ \theta $. Also the maximum
variation of $ c_{s}^{2} $ belongs to $ \Gamma = 0.75 $, from $ c_{s}^{2} \simeq 0.4 $ at $\theta = 0 $ (at nearly virial temperature)to about $ c_{s}^{2} = 0.12 $
at $ \theta = \pi /2 $. In this case, small value of $ \Gamma $  might be corresponding to the thin disk model with hot corona above the disk.
Most importantly, our results are totally in satisfactory with those presented in NY95a. In fact, we should mention that by using an analytical
solution for the case of no-wind self similar solutions, solving a system of complex ordinary deferential HD equations with two boundary condition
it is not necessary (readers referred to NY95a for more details).

\begin{figure}
\begin{center}
\includegraphics[width=90mm, angle=0]{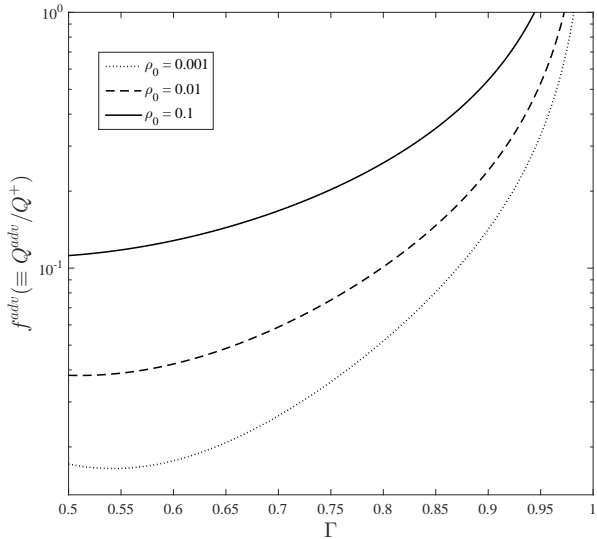}
\caption{Variation of the energy advection factor, $ f^{adv} (\equiv Q^{adv}/Q^{+}) $, with $ \Gamma $. The dotted, dashed and solid lines correspond to $ \rho_{0} = 0.001, 0.01, 0.1 $ respectively \textbf{with $ \alpha = 0.1 $.}}
\label{f_Gamma}
\end{center}
\end{figure}

As we explained before, the main purpose of this work is to check whether the advection parameter remains constant along the vertical direction or not. Fig. \ref{f_Gamma} represents the  variation of the energy advection factor, $ f^{adv}(\equiv Q^{adv}/Q^{+}) $, with $ \Gamma $ on the range $ 0.5\leq \Gamma< 1 $. In this figure  the values of the density varies over the range , i.e., $ \rho_{0} = 0.001, 0.01, 0.1 $. It is shown that $ f^{adv} $ increases with
increasing $ \Gamma $ and reaches to the unity for high value of $ \Gamma $ corresponding to fully advection case. In contrary to G15 conclusions,
this figure obviously demonstrates that in the case of no-outflow, i.e., $ v_{\theta} = 0 $,  there exists thermal equilibrium and therefore, advection
cooling can balance viscous heating. Also as it is seen, for  three typical values of the density on the rotation axis, the fully advection take place when
$ \Gamma\simeq 0.95 $. Therefore Fig. 1 together with this figure show when the energy equation is replaced with the polytropic equation of state, $ p = K \rho^{\Gamma} $, the fully advection case will be possible if constants $ K $ and $ \Gamma $ vary only in the range
$ 0 < K \leq 0.4 $ and $ 0.5 \leq \Gamma< 1 $ respectively.

\begin{figure}
\begin{center}
\includegraphics[width=90mm, angle=0]{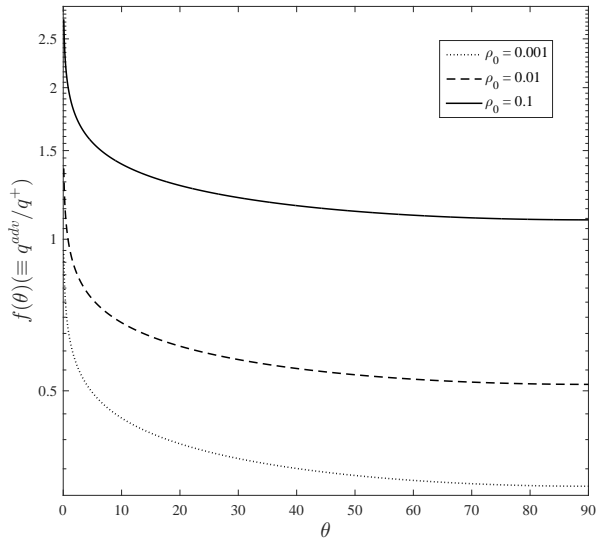}
\caption{Vertical profile of the energy advection factor, $ f(\theta) (\equiv q^{adv}/q^{+}) $. The dotted, dashed and solid lines correspond to $ \rho_{0} = 0.001, 0.01, 0.1 $ respectively. Here, $ \Gamma  = 0.95 $ and $ \alpha = 0.1 $.}
\label{f_theta}
\end{center}
\end{figure}

Finally, the vertical variation of advection factor, $ f(\theta) = q^{adv}/q^{vis} $, (the ratio of advection cooling to the viscose heating per unit volume) is plotted
in Fig. \ref{f_theta} for the same three values of the density in Fig. \ref{f_Gamma} with $ \Gamma = 0.95 $. It can be seen that the value
of the energy advection factor is not constant in the $ \theta $ direction and increase from equatorial plane towards the rotation axis.
In addition, the lower value of the polar axis density causes to the smaller value of $ f(\theta) $. It is also seen that the energy advection factor is
below unity, i.e., $ f(\theta) ‎‎\lesssim ‎1‎ $ for $ \rho =0.001 $. What's more, for the case with $ \rho = 0.01 $ the value of the energy
advection factor is first below unity an then become greater than one near the rotation axis. Furthermore, the larger value of the density close to the rotation axis,
i.e., $ \rho = 0.1 $, causes to the upper limit of $ f(\theta) $, i.e., $ f(\theta) \simeq‎‎ 2.8 $ at $\theta = 0 $ and more importantly this factor is
always above unity for all vertical range. Therefore unlike NY95a assumption which has been considered $ f = 1.0 $ throughout the angular direction, our result clearly
indicate that the advection parameter is a function of $ \theta $, i.e., $ f(\theta) $, and may exceed unity in some cases.

\section{Summery and conclusions}

In this paper, we tried to solve the hydrodynamic equations of optically thin ADAFs in spherical coordinate system $ (r, \theta, \phi) $ where the
steady accretion disk is considered to be symmetric with respect to the rotation axis as well as the equatorial plane. The central black hole gravity
is described as the Newtonian potential, since this form is more convenient in the self-similar solutions, $ \psi(r) = -(GM)/r $. In addition, instead
of using energy equation with constant value for the advection parameter, $ f $, following Gu et al. 2009 and G15, we adopted the polytropic relation in the vertical
direction as $ p = K \rho^{\Gamma} $. Compared with G15, we made three modifications. Firstly, the vertical range of the calculation is enhanced from
rotation axis $ \theta = 0^{\circ} $ to the equatorial plane of the disk, $ \theta = \pi/2 $. This change have been made since in optically thin ADAFs,
the disk is geometrically thick, i.e., $ H/R \lesssim 1 $. Secondly, following Narayan \& Yi 1995a to avoid singularity at the pole, the disk is taken to be symmetric with respect to this rotation axis. This change causes to find a relation between the value of constant $ K $ and the density at the rotation pole. Finally, the modified $ " \alpha " $ description of viscosity is adopted,
(see, e.g., Lovelace et al. 2009 for more details).

By the above mentioned modifications and following G15 methodology, we could find an analytical solution for optically thin ADAfs. The presented results showed that unlike G15, the radial velocity decreases towards the rotation axis for all $\Gamma$ values. In addition, $v_{\phi}$ becomes zero at rotation pole and this is due to the second modification. Furthermore, $c_{s}^2$, the same for NY95a, has a decreasing trend from $\theta =0$ (virial temperature) to $\theta = \pi /2$.

Besides in contrast to G15, our solution represents the existence of thermal equilibrium in vertical direction without outflow emanating. So advecting cooling can balance viscous heating effectively. Moreover, If a polytropic relation is used rather than the energy equation in the vertical direction for the fully advection, $ K $ and $ \Gamma $ should be only on the range $ 0 < K \leq 0.4 $ and $ 0.5 \leq\Gamma < 1 $, respectively. At last, the value
of the energy advection factor is not constant in $ \theta $ direction and increase from equatorial plane towards the rotation axis.

In spite of the simplicity of our model in viscosity and the disk itself, we think that the presented semi-analytical results
give us a better understanding of such a complicated system. It is good to note here that some modifications can be applied to ameliorate this study. Regarding to the radial self-similar approximation, it can not ensure us that this solution is definitely relevant to the real accretion flows. In addition, the Newtonian potential was
taking into account rather than Paczy\'{n}sky \& Wiita potential to avoid the general relativity effects in the innermost region of accretion disk.
Not only the $r\phi$ component of the viscous stress tensor, $ \tau_{r\phi} $, should be employed, but also the other components such as $ \tau_{\theta \phi} $ ineluctably should be taken into consideration. However, the advection factor $ f^{adv} $ in the energy equation was found to be function of the vertical direction, but this
parameter must be a function of radial and more significantly the mass accretion rate of the disk, i.e.,  $ \dot{m} $. To resolve the aformentioned remarks, considering $ v_{\theta} $ and also $ \tau_{\theta \phi} $ component of stress tensor may cause to have  more promising results in the future works.

\acknowledgements
The authors would like to thank Amin Mosallanezhad for his useful suggestions and discussions. We also would like to appreciate the referee for his/her thoughtful and constructive comments in the early version of the paper. This work was supported by Ferdowsi University of Mashhad under the grant 3/37875 (1394/04/03).

\end{document}